\documentclass{article}

    \PassOptionsToPackage{numbers}{natbib}

\usepackage[final]{new_in_ML}

\usepackage[utf8]{inputenc} 
\usepackage[T1]{fontenc}    
\usepackage{hyperref}       
\usepackage{url}            
\usepackage{booktabs}       
\usepackage{amsfonts}       
\usepackage{nicefrac}       
\usepackage{microtype}      
\usepackage{xcolor}         
\usepackage{amsmath}
\usepackage{graphicx}
\usepackage{relsize} 
\usepackage{caption}
\usepackage{tabularx}
\usepackage{multirow}
\usepackage{xspace}
\usepackage{makecell}
\usepackage{array} 
\usepackage{siunitx} 
\usepackage{pifont}
\usepackage{parskip}
\usepackage{enumitem}
\usepackage{threeparttable}

\usepackage{xspace}
\usepackage{lipsum}

\title{LightGCN: Evaluated and Enhanced\\}

\author{Milena Kapralova$^{1}$ \quad Luca Pantea$^{1}$ \quad Andrei Blahovici$^{1}$\\
$^1$ University of Amsterdam\\
\texttt{\{{milena.kapralova, luca.pantea, andrei.blahovici\}@student.uva.nl}}}

\begin{document}

\maketitle

\begin{abstract}
This paper analyses LightGCN \cite{he2020lightgcn} in the context of graph recommendation algorithms. Despite the initial design of Graph Convolutional Networks for graph classification, the non-linear operations are not always essential. LightGCN enables linear propagation of embeddings, enhancing performance. We reproduce the original findings, assess LightGCN's robustness on diverse datasets and metrics, and explore Graph Diffusion as an augmentation of signal propagation in LightGCN\footnote{Anonymised code repository: \url{https://anonymous.4open.science/r/LightGCN-1B84/README.md}}. 
\end{abstract}

\section{Introduction}

The Graph Convolution Network (GCN) has emerged as a powerful method in graph collaborative filtering recommendations. However, GCNs include many non-linear operations essential for collaborative filtering tasks. He et al. \cite{he2020lightgcn} proposed LightGCN, a model that removes non-linear activation functions and linear transformations from the Neural Graph Collaborative Filtering (NGCF) \cite{ndcg} architecture, linearly propagating user and item embeddings on the user-item bipartite graph. He et al. \cite{he2020lightgcn} empirically validated the new architectural choice by comparing it with NGCF and running an extensive ablation study over the model's hyperparameter space. This work has both reproducibility and extension goals. We verify that simpler GCNs achieve top performance through smoother embeddings. We further test LightGCN on five datasets, considering diversity and fairness, and enhance results using Graph Diffusion.


\section{Background}

\label{lightgcn}


The LightGCN model is based on the NGCF architecture, with the non-linear activation functions and linear projections removed. Thus, the aggregation function in the graph convolutions of LightGCN is:\vspace{-6mm}

\begin{align}
    &\mathbf{e}_u^{(k + 1)} = \sum_{i \in N_u} \frac{1}{\sqrt{|N_u|}\sqrt{|N_i|}} \mathbf{e}_i^{(k)}  
    &\mathbf{e}_i^{(k + 1)} = \sum_{u \in N_i} \frac{1}{\sqrt{|N_i|}\sqrt{|N_u|}} \mathbf{e}_u^{(k)}
\end{align}
\vspace{-5mm}

where $e_u^{(k)}$ and $e_i^{(k)}$ are embedding of user $u$ at layer $k$ and the embedding of item $i$ at layer $k$, respectively. LightGCN aggregates user and item embeddings from neighboursa at each layer, with only input embeddings being trainable. The final embeddings combine all layers' outputs. Predictions are based on the largest inner product between user and item embeddings: \vspace{-5mm}

\begin{equation}
\begin{aligned}
    \mathbf{e}_u &= \sum_{k=0}^K \alpha_k \mathbf{e}_u^{(k)} &\quad
    \mathbf{e}_i &= \sum_{k=0}^K \alpha_k \mathbf{e}_i^{(k)} &\quad
    \hat{y}_{ui} &= \mathbf{e}_u^\top \mathbf{e}_i
\end{aligned}
\end{equation}
\vspace{-4mm}

where $\alpha_k$ is the weight associated with the $k$-th layer embeddings. In He et al. \cite{he2020lightgcn}, $\forall k: \alpha_k = \frac{1}{K + 1}$.

\section{Method}
\label{APPNP}
Approximate Personalised Propagation of Neural Predictions (APPNP) is a GCN variant and an instance of graph diffusion convolution inspired by Personalised Page Rank (PPR). It derives from topic-sensitive PageRank approximated with power iteration \cite{gasteiger2022predict}, propagating the final embeddings: \vspace{-7 mm}

\begin{align}
    \textbf{Z}^{(0)} &= \textbf{E}^{(K)}, &
    \textbf{Z}^{(k+1)} &= \alpha \textbf{Z}^{(0)} + (1 - \alpha)\hat{\tilde{\textbf{A}}}\textbf{Z}^{(k)}, &
    \textbf{Z}^{(K)} &= \text{s} \left( \alpha \textbf{Z}^{(0)} + (1 - \alpha)\hat{\tilde{\textbf{A}}}\textbf{Z}^{(K-1)}  \right)
\end{align}
\vspace{-5mm}

The final embedding matrix $\textbf{E}^{(K)}$ serves as the starting vector and teleport set with $K$ denoting power iteration steps. This method maintains graph sparsity without needing extra training parameters and prevents oversmoothing due to its teleport design. The teleport probability, $\alpha$, adjusts the neighbourhood size. While Gasteiger et al. \cite{gasteiger2022predict} found optimal alphas between [.05, .2], our grid search on the CiteULike dataset identified the best $\alpha$ as .1. For other datasets, we begin with this alpha, adjusting slightly to optimize test set performance.


\section{Experimental setup}
\label{sec:setup}

\label{sec:datasets}
\textit{Datasets.} He et al. \cite{he2020lightgcn} use the datasets Gowalla, Yelp2018, and Amazon-Book. To see how LightGCN performs in different domains and dataset sizes, we use five additional datasets (Table \ref{tab:statistics}).

\vspace{-3mm}
\begin{table}[ht]
\centering
\caption{Statistics of the experimented datasets}
\label{tab:statistics}
\resizebox{1\textwidth}{!}{
\begin{tabular}{lcccccccc}
\toprule
\textbf{Dataset} & \textbf{Gowalla} & \textbf{Yelp2018} & \textbf{A-Book} & \textbf{CiteULike*} & \textbf{A-Movies*} & \textbf{A-Electro*} & \textbf{A-CDs*} & \textbf{A-Beauty*} \\
\midrule
\textbf{User \#} & 29,858 & 31,668 & 52,643 & 3,276 & 44,438 & 1,434 & 43,168 & 7,068 \\
\textbf{Item \#} & 40,981 & 38,048 & 91,599 & 16,807 & 25,046 & 1,522 & 35,647 & 3,569 \\
\textbf{Inter. \#} & 1,027,370 & 1,561,406 & 2,984,108 & 178,062 & 1,070,860 & 35,931 & 777,426 & 79,506 \\
\textbf{Density} & 0.00084 & 0.00130 & 0.00062 & 0.00323 & 0.00096 & 0.01645 & 0.00051 & 0.00315 \\
\bottomrule
\end{tabular}
}
\begin{tablenotes}
\item[a] Note: ``A'' indicates Amazon. *Additional datasets not implemented in He et al.\cite{he2020lightgcn}
\end{tablenotes}
\end{table}

\vspace{-3mm}
\label{sec:metrics}
\textit{Metrics.} We assess NDCG, recall, precision, diversity and fairness. We calculate \textit{diversity} by the ILD \cite{cen2020controllable}. Users who make repeat purchases receive better recommendations than those exploring new items or with few interactions \cite{10.1145/3587153}. To calculate \textit{fairness}, we bin users by the number of interacted items. Similar recommendation performance across bins implies more fairness.

\textit{Implementation details}. We follow He et al. \cite{he2020lightgcn} in training the models for 1000 epochs (for diffusion experiments, we use 600 epochs), using Adam optimizer ($lr=.001$), the BPR loss and $\lambda= .0001$.


\section{Replication study}
\label{sec:replication-study}

\textbf{Claim 1: LightGCN Surpasses SOTA GCN Recommenders - \textit{Correct}}


Table \ref{tab:tab1} indicates the replicated LightGCN underperforms compared to the original, with Gowalla showing improvement with more layers, while Yelp2018 and Amazon-Book have varied outcomes. Table \ref{tab:tab2} examines the 3-layer LightGCN with different normalizations. Consistent with the original paper, square root normalization on both sides is best, while altering one side diminishes results. Using $L_1$ normalization symmetrically reduces performance for Gowalla and Yelp2018 but enhances it for Amazon-Book. Despite variations, the replicated values surpass NGCF.

\vspace{-5mm}
\begin{table}[ht]
\centering
\caption{Replicated ($\mathbf{Rep}$.) LightGCN. The relative difference to the original results is in \%.}
\vspace{3mm}
\resizebox{1\textwidth}{!}{%
\begin{tabular}{ccccccccc}
\toprule
\multicolumn{2}{c}{\textbf{Dataset}} & \multicolumn{2}{c}{\textbf{Gowalla}} & \multicolumn{2}{c}{\textbf{Yelp2018}} & \multicolumn{2}{c}{\textbf{Amazon-Book}} \\
\cmidrule(r){1-2} \cmidrule(lr){3-4} \cmidrule(lr){5-6} \cmidrule(lr){7-8}
\textbf{Layer \#} & \textbf{Method} & \textbf{Recall@20} & \textbf{NDCG@20} & \textbf{Recall@20} & \textbf{NDCG@20} & \textbf{Recall@20} & \textbf{NDCG@20} \\
\midrule
\multirow{1}{*}{\textbf{1 Layer}} 
& LightGCN ($\mathbf{Rep}$.) & 0.1666 (\textcolor{red}{-5.08\%}) & 0.1399 (\textcolor{red}{-6.34\%}) & 0.0557 (\textcolor{red}{-13.87\%}) & 0.0452 (\textcolor{red}{-12.14\%}) & 0.0383 (\textcolor{red}{-0.26\%}) & 0.0293 (\textcolor{red}{-1.64\%}) \\
\midrule
\multirow{1}{*}{\textbf{2 Layers}} 
& LightGCN ($\mathbf{Rep}$.) & 0.1774 (\textcolor{red}{-0.17\%}) & 0.1512 (\textcolor{red}{-0.80\%}) & 0.0613 (\textcolor{red}{-1.44\%}) & 0.0503 (\textcolor{red}{-0.20\%}) & 0.0409 (\textcolor{red}{-0.97\%}) & 0.0316 (\textcolor{green}{+0.32\%}) \\
\midrule
\multirow{1}{*}{\textbf{3 Layers}} 
& LightGCN ($\mathbf{Rep}$.) & 0.1807 (\textcolor{red}{-0.88\%}) & 0.1537 (\textcolor{red}{-1.07\%}) & 0.0643 (\textcolor{green}{+0.63\%}) & 0.0528 (\textcolor{green}{+0.62\%}) & 0.0416 (\textcolor{green}{+1.46\%}) & 0.0322 (\textcolor{green}{+2.13\%}) \\
\midrule
\multirow{1}{*}{\textbf{4 Layers}} 
& LightGCN ($\mathbf{Rep}$.) & 0.1806 (\textcolor{red}{-1.31\%}) & 0.1528 (\textcolor{red}{-1.20\%}) & 0.0653 (\textcolor{green}{+0.62\%}) & 0.0537 (\textcolor{green}{+1.28\%}) & 0.0414 (\textcolor{green}{+1.79\%}) & 0.0319 (\textcolor{green}{+2.12\%}) \\
\bottomrule
\end{tabular}%
}
\label{tab:tab1}
\end{table}

\textbf{Claim 2: Evaluating Embedding Smoothing in LightGCN's Performance Gain - \textit{Unreproducible}}

The step needs $O(M+N)^2$ operations ($M$ - \#users, $N$ - \#items); parallelization isn't viable due to set intersection. With an estimated 1507-hour runtime, we didn't replicate this study aspect.

\vspace{-5mm}
\begin{table}[ht]
\centering
\caption{3-layer LightGCN with different normalization schemes in graph convolution.}
\vspace{2mm}
\resizebox{1\textwidth}{!}{
\begin{tabular}{lccccccc}
\toprule
\multirow{2}{*}{{\textbf{Method}}} & \multicolumn{2}{c}{\textbf{Gowalla}} & \multicolumn{2}{c}{\textbf{Yelp2018}} & \multicolumn{2}{c}{\textbf{Amazon-Book}} & \multirow{2}{*}{\shortstack{\textbf{Replication Rel.}\\ \textbf{Diff. Interval}}} \\
\cmidrule(lr){2-3} \cmidrule(lr){4-5} \cmidrule(lr){6-7}
 & \textbf{Recall@20} & \textbf{NDCG@20} & \textbf{Recall@20} & \textbf{NDCG@20} & \textbf{Recall@20} & \textbf{NDCG@20} & \\
\midrule
LightGCN-$L_1$-L & 0.1700 & 0.1392 & 0.0623 & 0.0507 & 0.0424 & 0.0324 & $[-1.55\%, \ 1.25\%]$ \\
LightGCN-$L_1$-R & 0.1517 & 0.1223 & 0.0553 & 0.0445 & 0.0321 & 0.0246 & $[-9.28\%, \ -3.87\%]$ \\
LightGCN-$L_1$ & 0.1564 & 0.1300 & 0.0570 & 0.0462 & 0.0360 & 0.0276 & $[-1.64\%, \ 0.36\%]$ \\
LightGCN-L & 0.1767 & 0.1501 & 0.0642 & 0.0524 & 0.04102 & 0.03216 & $[-0.07\%, 14.01\%]$ \\
LightGCN-R & 0.173 & 0.1436 & 0.06189 & 0.0499 & 0.03695 & 0.02817 & $[-0.46\%, 24.44\%]$ \\
LightGCN & 0.1807 & 0.1537 & 0.0643 & 0.0528 & 0.0416 & 0.0322 & $[-1.25\%, \ 2.22\%]$ \\
\bottomrule
\end{tabular}%
}
\begin{tablenotes}
\item[a] Notation: $-L$ and $-R$ the side norm used, $-L_1$ and $L_2$ indicate the norm used.
\end{tablenotes}
\label{tab:tab2}
\end{table}

\section{Extended work}
\label{sec:extended}





\textit{Additional datasets} In examining various datasets, the 4-layer configuration of the LightGCN variant consistently outperformed the other configurations. For the \textbf{CiteULike} dataset, the 4-layer setup improved performance by roughly 44-50\% over the 1-layer configuration. In the \textbf{Amazon-Movies} dataset, the enhancement was around 30\%. \textbf{Amazon-Electronics} showcased an increase of 31.8-44.4\%, while \textbf{Amazon-CDs} and \textbf{Amazon-Beauty} revealed improvements of 11.5-16.3\%. Notably, datasets with reduced density, e.g. Amazon-Electronics, tend to yield weaker performance outcomes compared to denser datasets.

\vspace{-1mm}

\begin{figure}[ht]
  \centering
  \begin{tabular}{@{}ccc@{}}
    \includegraphics[width=0.31\textwidth]{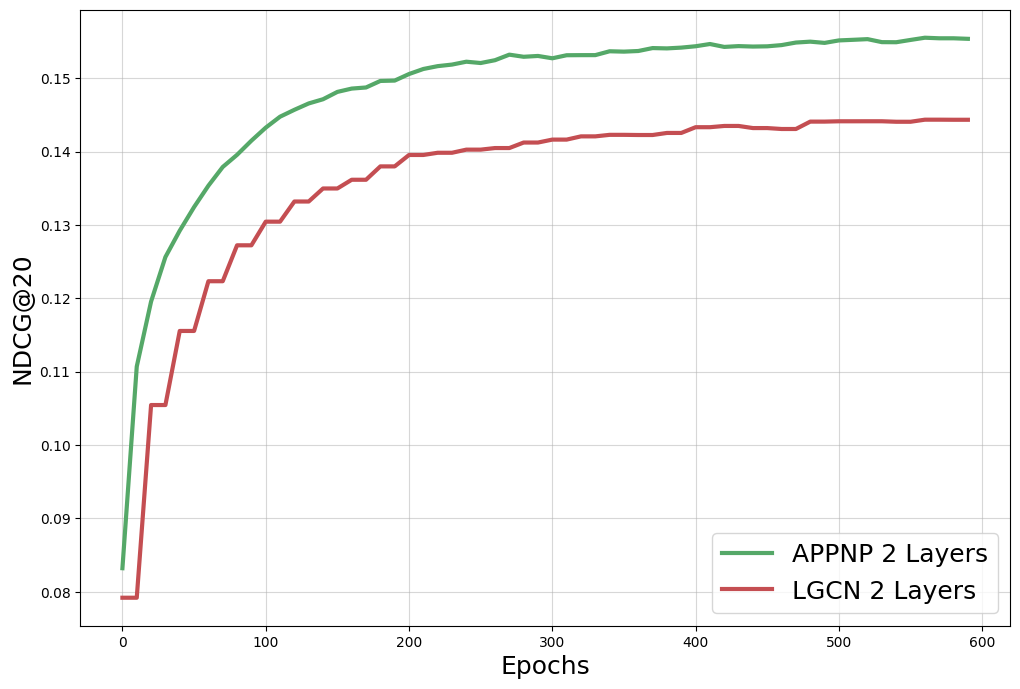} &
    \includegraphics[width=0.31\textwidth]{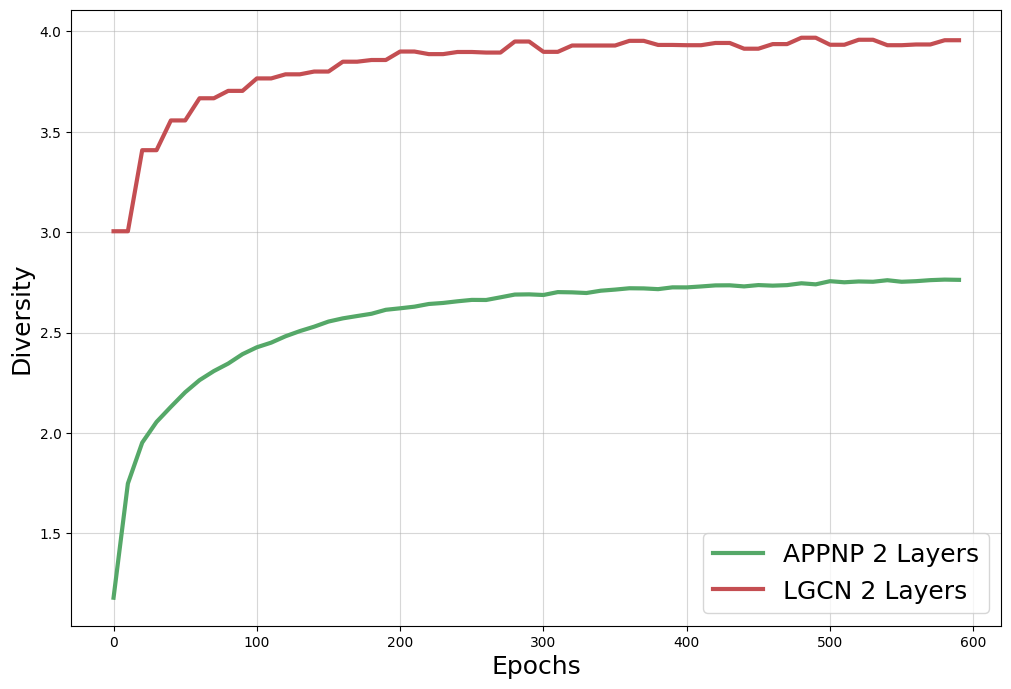} &
    \includegraphics[width=0.31\linewidth]{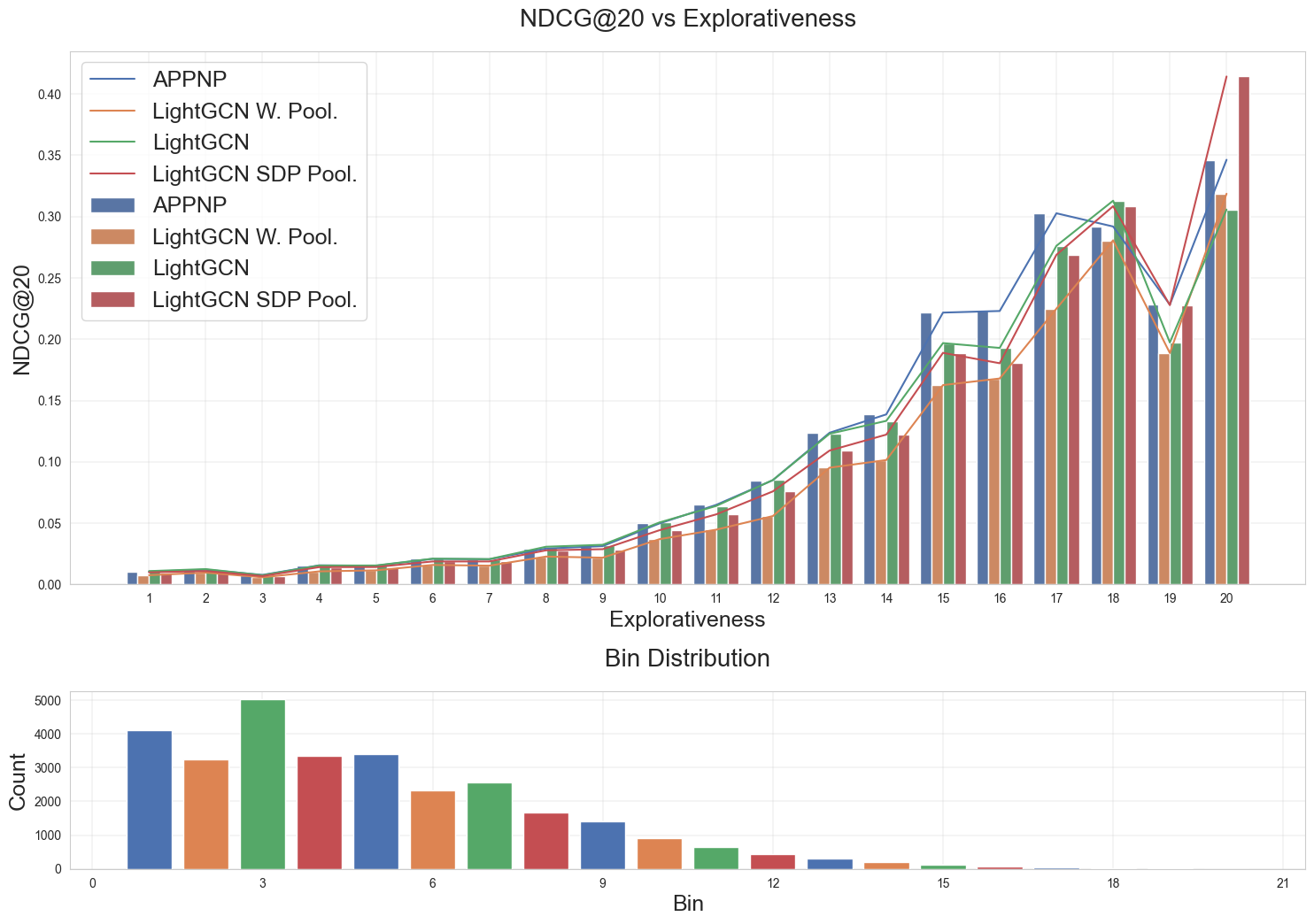} 
  \end{tabular}
  
  \caption{NDCG, diversity and fairness with APPNP and LightGCN, on the Gowalla dataset.}
  \label{fig:diversity}
\end{figure}

\textit{Additional metrics.} In Figure \ref{fig:diversity}, we compare LightGCN and APPNP on NDCG@20 and \textbf{diversity}. APPNP outperforms in NDCG@20 but not in diversity, stressing the need for comprehensive evaluation. Limited propagation iterations in APPNP may hinder diverse information diffusion. For \textbf{fairness} (Figure \ref{fig:diversity}), both models favour users with more interaction data, with APPNP being less biased.


\vspace{-2mm}

\begin{figure}[ht]
  \centering
  \begin{tabular}{@{}ccc@{}}
    \includegraphics[width=0.4\textwidth]{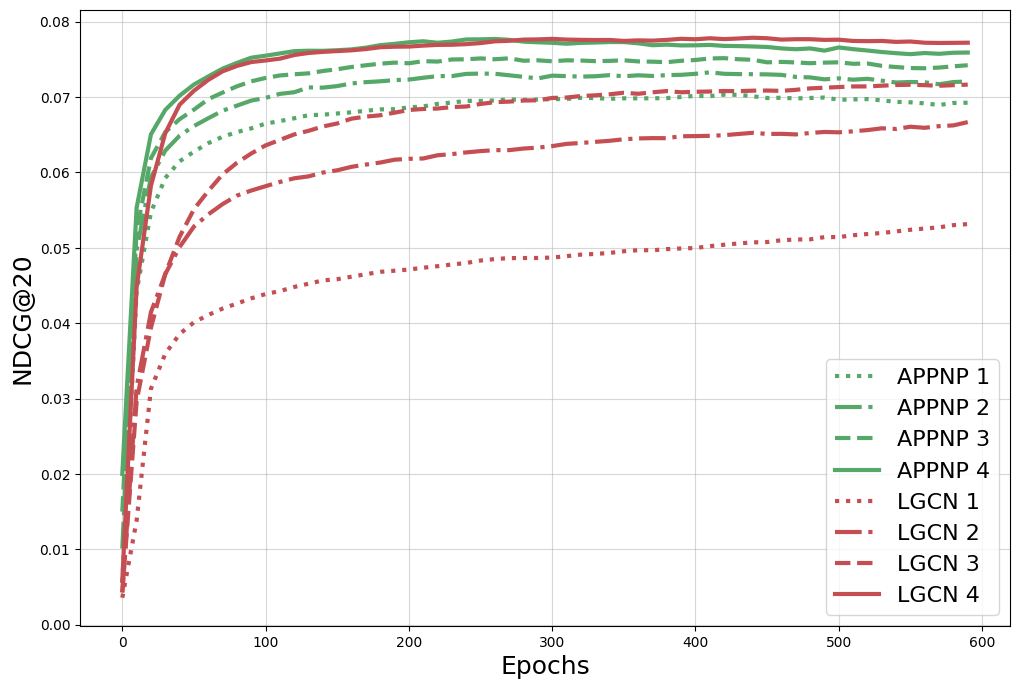} &
    \includegraphics[width=0.4\textwidth]{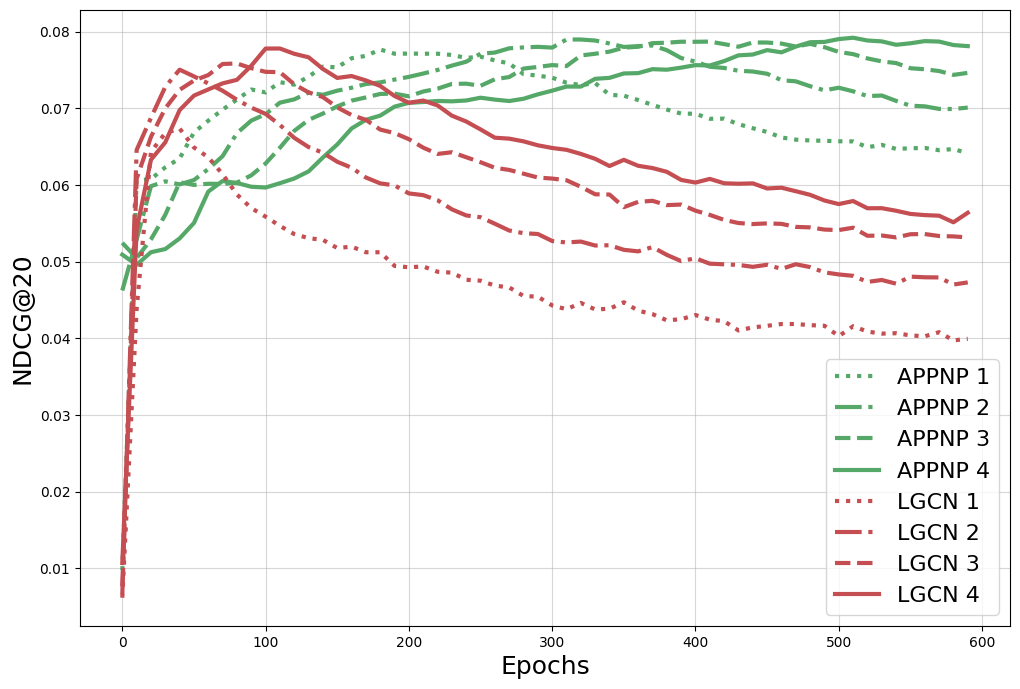} &
  \end{tabular}
  \caption{Model performance with (APPNP) and without (LGCN) propagation of embeddings, on diverse datasets. For both APPNP models, number of propagation steps=10 $\alpha$=0.1}
  \label{fig:diffusion}
\end{figure}

\textit{Graph diffusion.} Using propagation on LightGCN embeddings enhances early training stability (Figure \ref{fig:diffusion}) without significantly improving maximum performance compared to LightGCN alone. Notably, LightGCN's performance declines during training on the Amazon-Electronics dataset, in contrast to APPNP, which maintains stability. Similar trends are observed across other datasets.


\vspace{-5mm}

\section{Discussion and Conclusion}

This study mostly replicated He et al.'s \cite{he2020lightgcn} findings, even though time constraints limited full replication. LightGCN excels with high-interaction users but is generally outperformed by diffusion-augmented embeddings. However, combining LightGCN with APPNP can prolong training and require parameter tuning. Future work can consider approaches like Dual LightGCN for graded interaction predictions, integrating graph attention networks for complex relationship capture and understanding how dataset properties such as sparsity and graph size impact diffusion parameters.


\bibliographystyle{plain}
\bibliography{references} 
\end{document}